\documentclass[11pt,a4paper,notitlepage,onecolumn]{article}
\usepackage[a4paper,width=170mm,top=25mm,bottom=25mm]{geometry}
\RequirePackage{latex/atlaslatexpath}
\usepackage[subcaption]{atlaspackage}
\usepackage{atlasbiblatex}
\usepackage{atlasphysics}
\graphicspath{{logos/}{figures/}}

\title{Highlights on top-quark physics with \\ the ATLAS experiment at the LHC}
\author{Benedikt Gocke, on behalf of the ATLAS Collaboration}
\date{}

\begin{document}
	
	\maketitle
	\begin{abstract}
	The large top-quark samples collected with the ATLAS experiment at the LHC have yielded measurements of the inclusive \ttbar production cross section of unprecedented precision and differential measurements in new kinematic regimes. 
	They have also enabled new measurements of top-quark properties that were previously inaccessible, enabled the observation of many rare top-quark production processes predicted by the Standard Model and boosted searches for flavour-changing-neutral-current interactions of the top-quark, that are heavily suppressed in the SM. 
	In this contribution the highlights of the ATLAS top-quark physics program are presented, as well as projections of the expected sensitivity after the High Luminosity phase of the LHC.\\\\
	Talk presented at the International Workshop on Future Linear Colliders (LCWS 2023), 15-19 May 2023. C23-05-15.3.
	\end{abstract}
	\let\footnoterule\relax
	\let\thefootnote\relax
	\footnotetext{Copyright 2023 CERN for the benefit of the ATLAS Collaboration. Reproduction of this article or parts of it is allowed as specified in the CC-BY-4.0 license}

	\clearpage
	\section{Introduction}
	\label{sec:intro}
	
	The top-quark is the heaviest knwon elementary particle in the Standard model (SM).
	Due to its large mass, the top-quark decays before it hadronises and thus, allows in principle to directly measure its properties.
	Further, top-quark processes are crucial for searches for Beyond the Standard Model (BSM) processes, as BSM contributions could alter the probability of top-quark-involved processes.
	Also, top-quark processes, especially with associated particles (e.g. bosons), are important background processes not only to BSM searches but for measurements of Higgs-Boson procsses.   
	Measuring these processes allows to test the SM and BSM theories.
	Therefore, the top-quark properties, e.g. mass, spin or couplings, and production modes via strong and electroweak interaction need to be measured and known with high precision.\\
	The top-quark working group in the ATLAS experiment presents some of the recent highlights of top-quark measurements.
	Using the full Large Hadron Collider (LHC) Run-2 dataset at a centre-of-mass energy of $\sqrt{s}=13$ TeV amounting to a measured integrated luminosity of $\mathcal{L}=140$ fb$^{-1}$~~\cite{DAPR-2021-01}, allows for the most precise measurements to date.
	Top-quark properties measurement are shown, as well as inclusive and differential cross-sections measurements.
	For all measurements, it is crucial, that theory predictions for simulated Monte Carlo (MC) events are well modelled to obtain precise results.
	Today, the uncertainties in signal and background processes are often limiting the overall precision of the measurements.
	Using differential cross-section measurements help understanding and comparing different theory predictions.
	Further, first observations of rare processes are highlighted in the following selected results.
	
	\section{Highlights on top-quark analyses}
	\label{sec:highlights}
	
	\subsection{Inclusive and differential $\boldsymbol{t\bar{t}}$ production cross-section measurement}
	\label{sec:tchannel}
\begin{figure}
		\centering
		\includegraphics[scale=0.35]{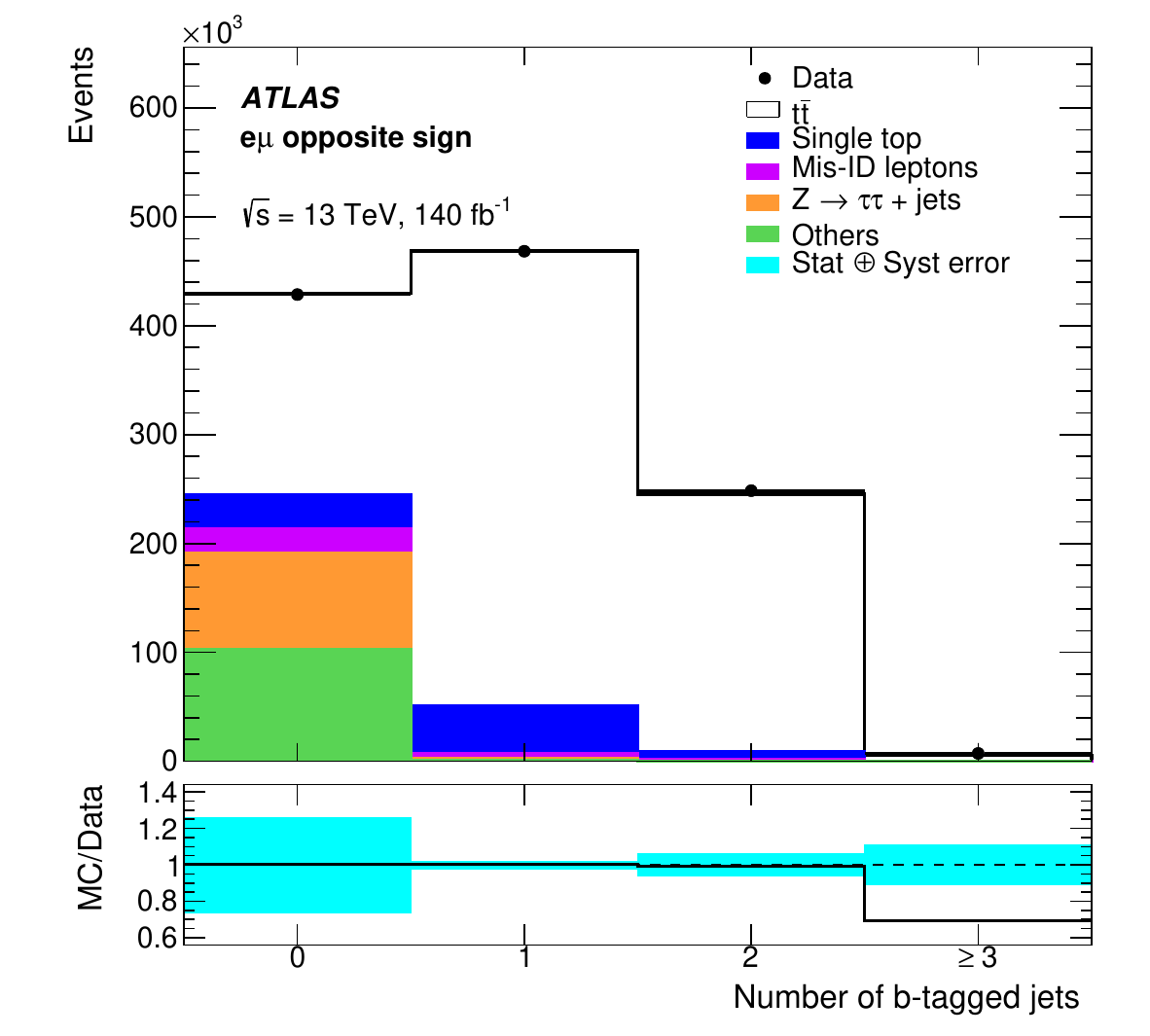}
		\caption{Distribution of the number of $b$-tagged jets in selected opposite-sign $e\mu$ events. The coloured distributions show the breakdown of the predicted background contributions from single top-quarks ($Wt$ and $t$-channel), misidentified leptons, $Z (\rightarrow \tau\tau)$ + jets and other sources of background (diboson, $t\bar{t}W$, $t\bar{t}Z$, and $t\bar{t}H$). The bottom panel shows the ratio of the prediction to the data with an uncertainty band covering both the statistical and systematic uncertainties, except for $t\bar{t}$ generator uncertainties.~\cite{TOPQ-2018-26}}
		\label{fig:tchannel/nbjets}
	\end{figure}

\begin{figure}
		\centering
		\begin{subfigure}{0.48\textwidth}
			\includegraphics[scale=0.35]{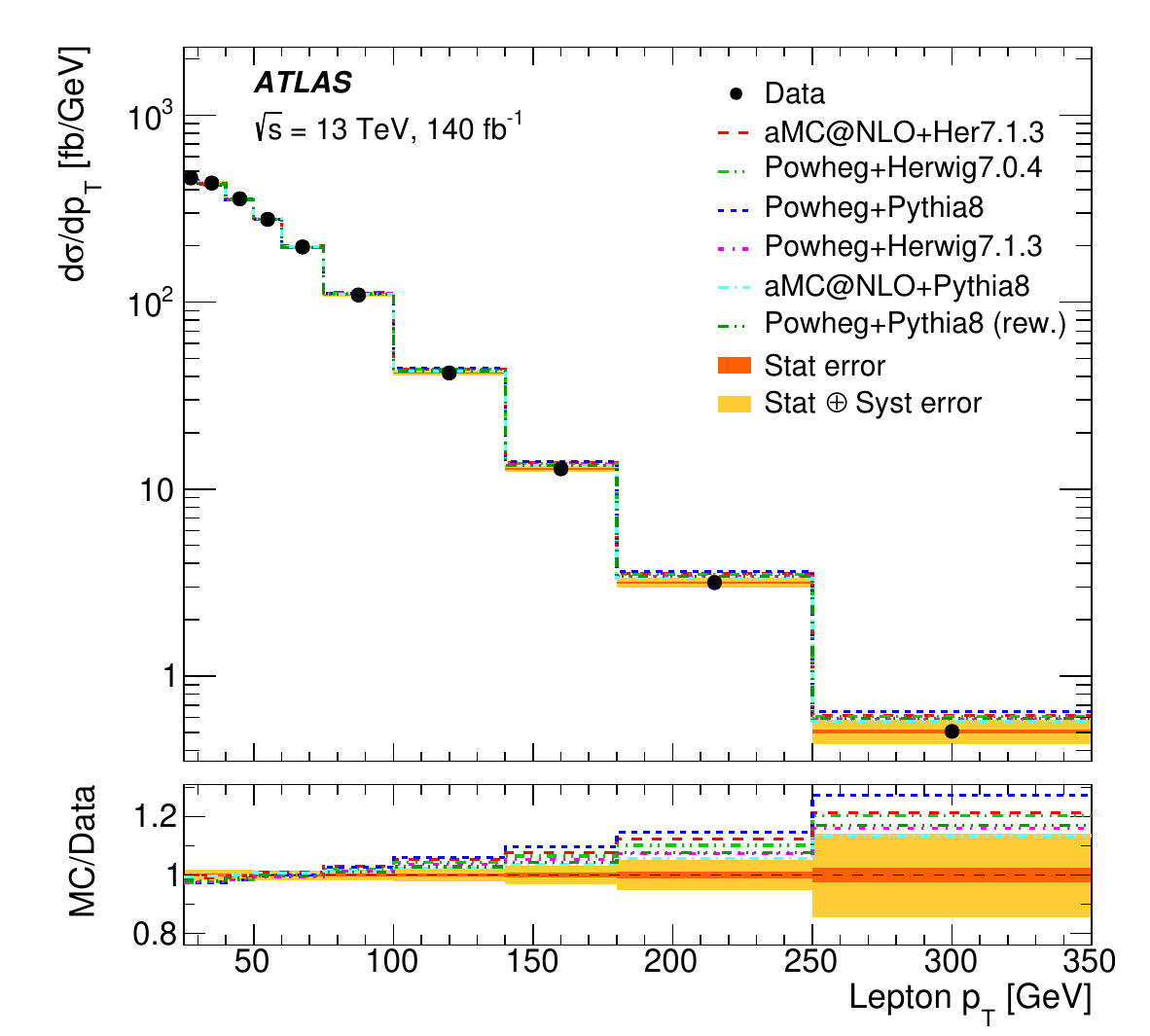}
			\caption{}
		\end{subfigure}
		\hfill
		\begin{subfigure}{0.48\textwidth}
			\hspace*{-1.6cm}
			\includegraphics[scale=0.5]{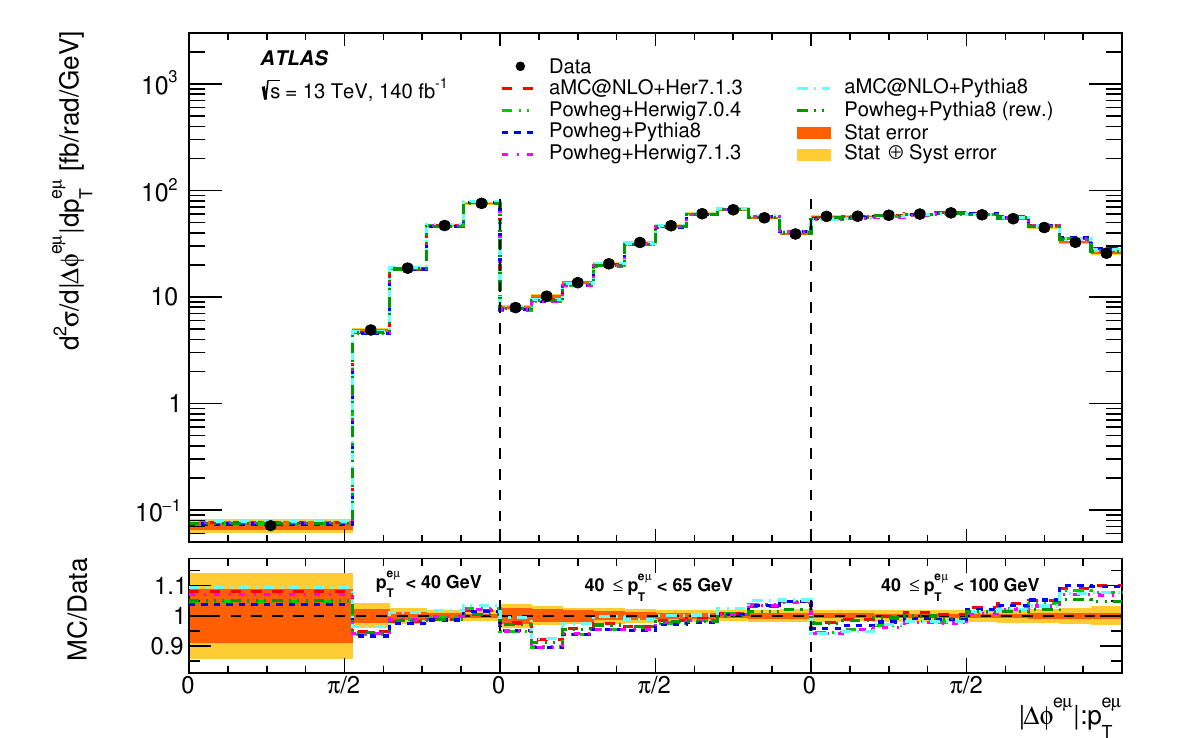}
			\caption{}
		\end{subfigure}
		\caption{(a) Absolute differential cross-sections as a function of $\pt^{\ell}$ with statistical (orange) and statistical plus systematic uncertainties (yellow) and (b) differential cross-sections as a function of $\lvert \Delta\phi(e,\mu)\rvert$ in bins of $\pt^{e\mu}$. The data points are shown as black dots and are placed at the centre of each bin. The results are compared with the predictions from different Monte Carlo generators normalised to the Top++ NNLO+NNLL prediction: the baseline \Powheg+\Pythia 8.230 $t\bar{t}$ sample (blue), \AMCatNLO+\Herwig 7.1.3 (red), \Powheg+\Herwig 7.0.4 (green), \Powheg+\Herwig 7.1.3 (purple), \AMCatNLO+\Pythia 8.230 (cyan) and \Powheg+\Pythia 8.230 rew. (dark green), which refers to \Powheg+\Pythia 8.230 reweighted according to the top-quark $\pt$. The lower panel shows the ratios of the predictions to data, with the bands indicating the statistical and systematic uncertainties. The last bin in also contains overflow events.~\cite{TOPQ-2018-26}}
		\label{fig:tchannel/diff}
	\end{figure}

	The measurement of the inclusive and differential $t\bar{t}$ production cross-section was done using the full Run-2 dataset obtained with the ATLAS detector at $\sqrt{s} = 13$ TeV accounting to an integrated luminosity of $\mathcal{L} = 140$ fb$^{-1}$~\cite{TOPQ-2018-26}. 
	Events are selected with exactly one electron and exactly one muon and either exactly one or two $b$-tagged jets.
	These selection criteria ensure are minimal level of background.
	This is shown in Figure~\ref{fig:tchannel/nbjets}, where events with one electron and one muon in the final state are split into number of $b$-tagged jets.
	For events with one or two $b$-tagged jets, the distribution is very pure in signal events.
	Misidentified leptons and $Z\rightarrow\tau\tau$+jets background are calculated using data-driven methods.
	Both, the inclusive and the differential cross-section measurements use a log-likelihood fit to the number of selected events $N$.
	The two equations used in the likelihood-formula are shown in Equations~\eqref{eq:tchannel/likelihood1},\eqref{eq:tchannel/likelihood2}:
	\begin{align}
	N^{i}_{\mathrm{1}} &= \mathcal{L}\sigma^{i}_{t\bar{t}}G^{i}_{e\mu}2\epsilon_{b}^{i} \left(1-\epsilon_{b}^{i}C^{i}_{b}\right) + N^{i}_{1,\mathrm{bkg}}\label{eq:tchannel/likelihood1}\\
	N^{i}_{\mathrm{1}} &= \mathcal{L}\sigma^{i}_{t\bar{t}}G^{i}_{e\mu}\left(\epsilon_{b}^{i}\right)^{2}C^{i}_{b} + N^{i}_{1,\mathrm{bkg}} \label{eq:tchannel/likelihood2}
	\end{align}
	For the differential measurement, the likelihood fit is done in each bin $i$, whereas for the inclusive measurement, the fit is performed in two inclusive bins. 
	Performing the fit allows a simultaneous determination of the cross-section $\sigma^{i}_{t\bar{t}}$ and the combined jet selection and $b$-tagging efficiency $\epsilon_b^{i}$.
	The reconstruction efficiency, $G^{i}_{e\mu}$, is defined as the number of selected lepton pairs in the $t\bar{t}$ sample, which are reconstructed in bin $i$, divided by the total number of lepton pairs generated in bin $i$.
	The $b$-tagging correlation coefficient, $C^{i}_{b}$, corrects the probability of tagging the second jet after having tagged the first one.
	It is also determined by simulation and is found to be close to unity.
	The inclusive cross-section is measured to be
	\begin{equation}
	\sigma_{t\bar{t}} = 829 \pm 1 (\mathrm{stat}) \pm 13 (\mathrm{syst}) \pm 8 (\mathrm{lumi}) \pm 2 (\mathrm{beam})\, \mathrm{pb}
	\end{equation}
	The newest luminosity measurement of ATLAS~\cite{DAPR-2021-01} allows for an incredible precision, as it lowers the relative uncertainty below $2\%$ for this measurement.
	The overall uncertainty of below 2\% shows the strength of the LHC as a precision machine.
	For the differential cross-section measurement eight single lepton kinematic variables ($\pt^{\ell}$,$\lvert\eta^{\ell}\rvert$,$m^{e,\mu}$, $\pt^{e,\mu}$, $\lvert y^{e,\mu} \rvert$, $E^e + E^\mu$, $\pt^\mu + \pt^e$, $\lvert \Delta\phi^{e,\mu}$) were chosen and four double differential distributions ($\lvert y^{e,\mu}\rvert$ in five bins of $m^{e,\mu}$,$\lvert \Delta\phi^{e,\mu} \rvert$ in five bins of $m^{e,\mu}$,$\lvert \Delta\phi^{e,\mu} \rvert$ in three bins of $\pt^{e,\mu}$,$\lvert \Delta\phi^{e,\mu} \rvert$ in five bins of $E^e + E^\mu$).
	Again, this measurement benefits from exploiting the full Run-2 dataset, as a wider range in distributions is possible, as well as a finer granularity for the chosen bin sizes.
	Two example distributions are shown in Figure~\ref{fig:tchannel/diff}.
	The results show that a more refined theoretical modelling is needed, as discrepancies with predictions from several next-to-leading order (NLO) event-generators are visible in the distributions. 
	
	\subsection{$t\bar{t}$ and $Z$-boson cross sections and their ratio at $\sqrt{s}=13.6$ TeV}
	\label{sec:ttbarRun3}
\begin{figure}
		\centering
		\begin{subfigure}{0.48\textwidth}
			\includegraphics[scale=0.35]{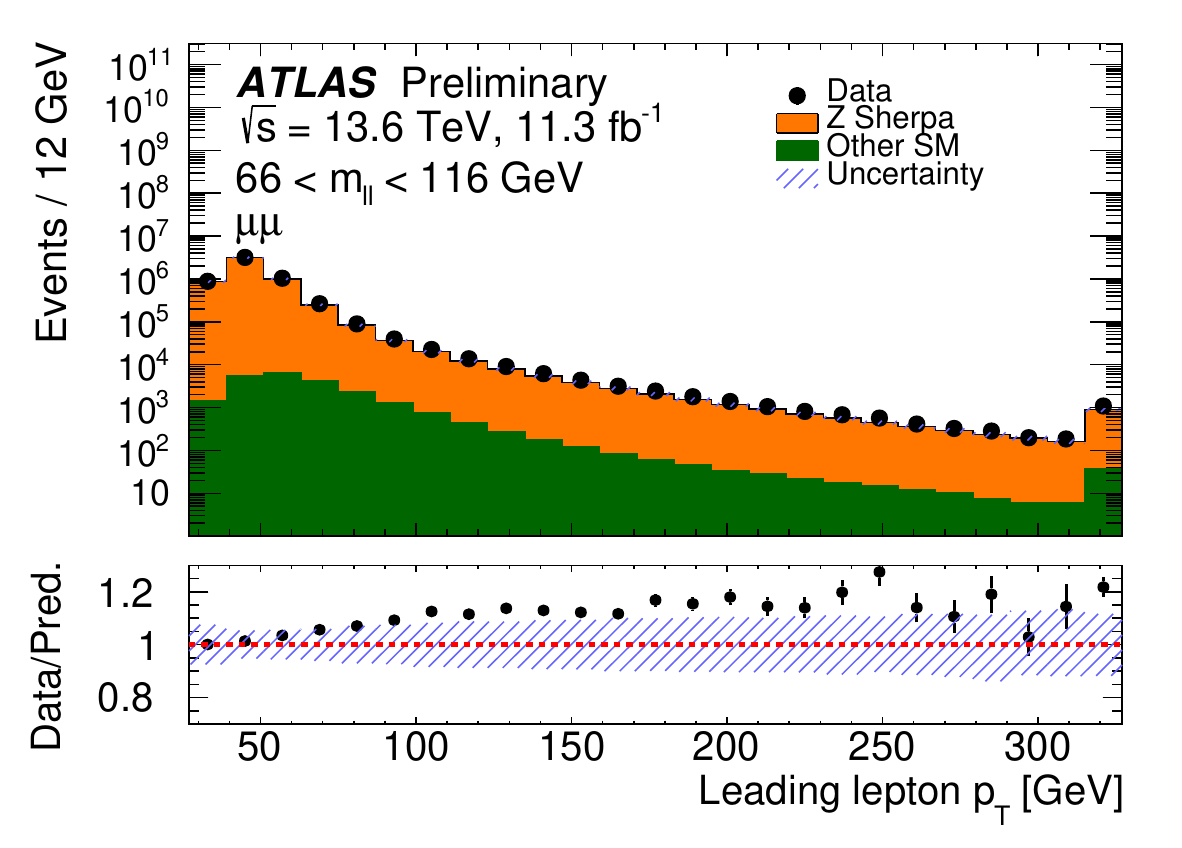}
			\caption{}
		\end{subfigure}
		\begin{subfigure}{0.48\textwidth}
			\includegraphics[scale=0.35]{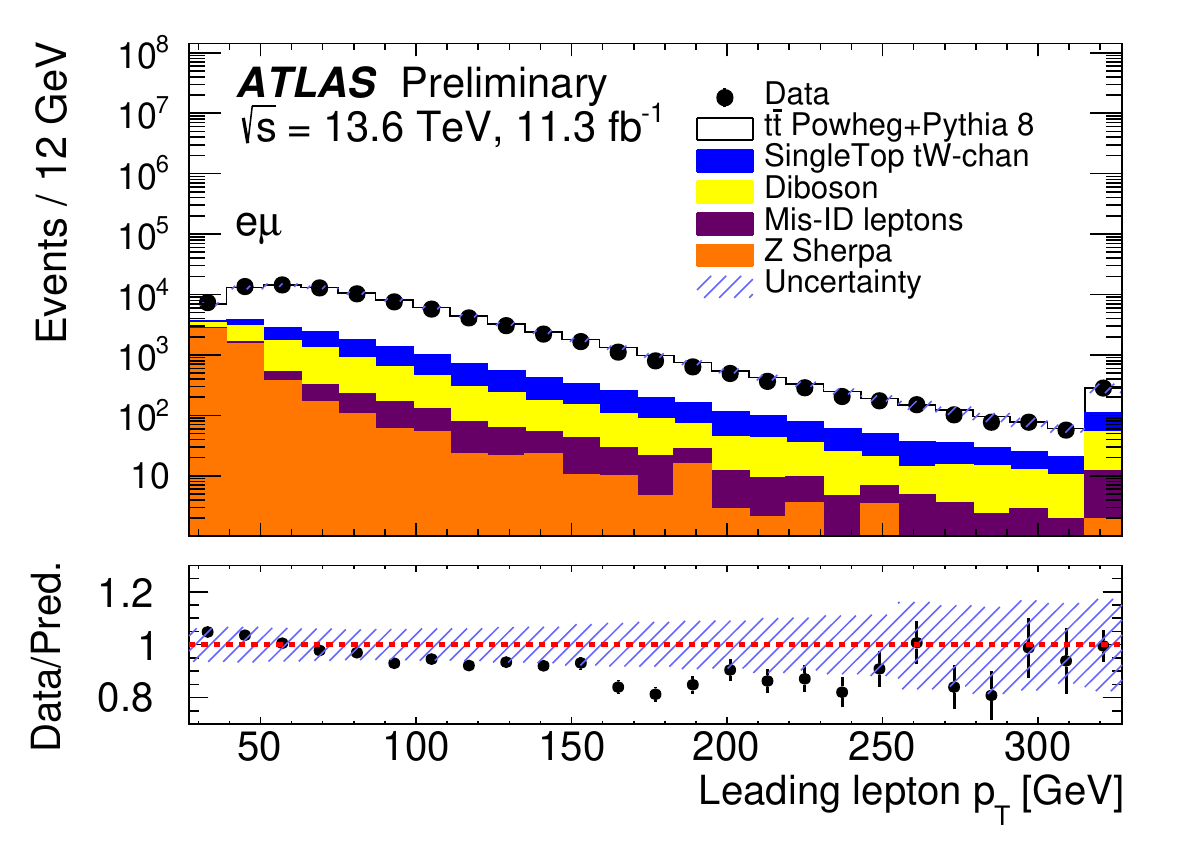}
			\caption{}
		\end{subfigure}
		\caption{Comparison of observed data and predictions for the \pt of the leading lepton in (a) the $\mu\mu$ channel and (b) the \pt of the leading lepton in the $e\mu$ channel, in the Run-3 \ttbar and $Z$-boson cross-section measurement. The expected yields are calculated by normalising the MC prediction using the cross-section for each process and the estimate of the data integrated luminosity. The "Mis-ID" label represents fake and non-prompt leptons. The hashed band represents the total uncertainty. The bottom panel shows the ratio of data to prediction. The rightmost bins contain the overflow events.~\cite{ATLAS-CONF-2023-006}}
		\label{fig:ttbarRun3/dataMC}
	\end{figure}

\begin{figure}
		\centering
		\begin{subfigure}{0.48\textwidth}
			\includegraphics[scale=0.35]{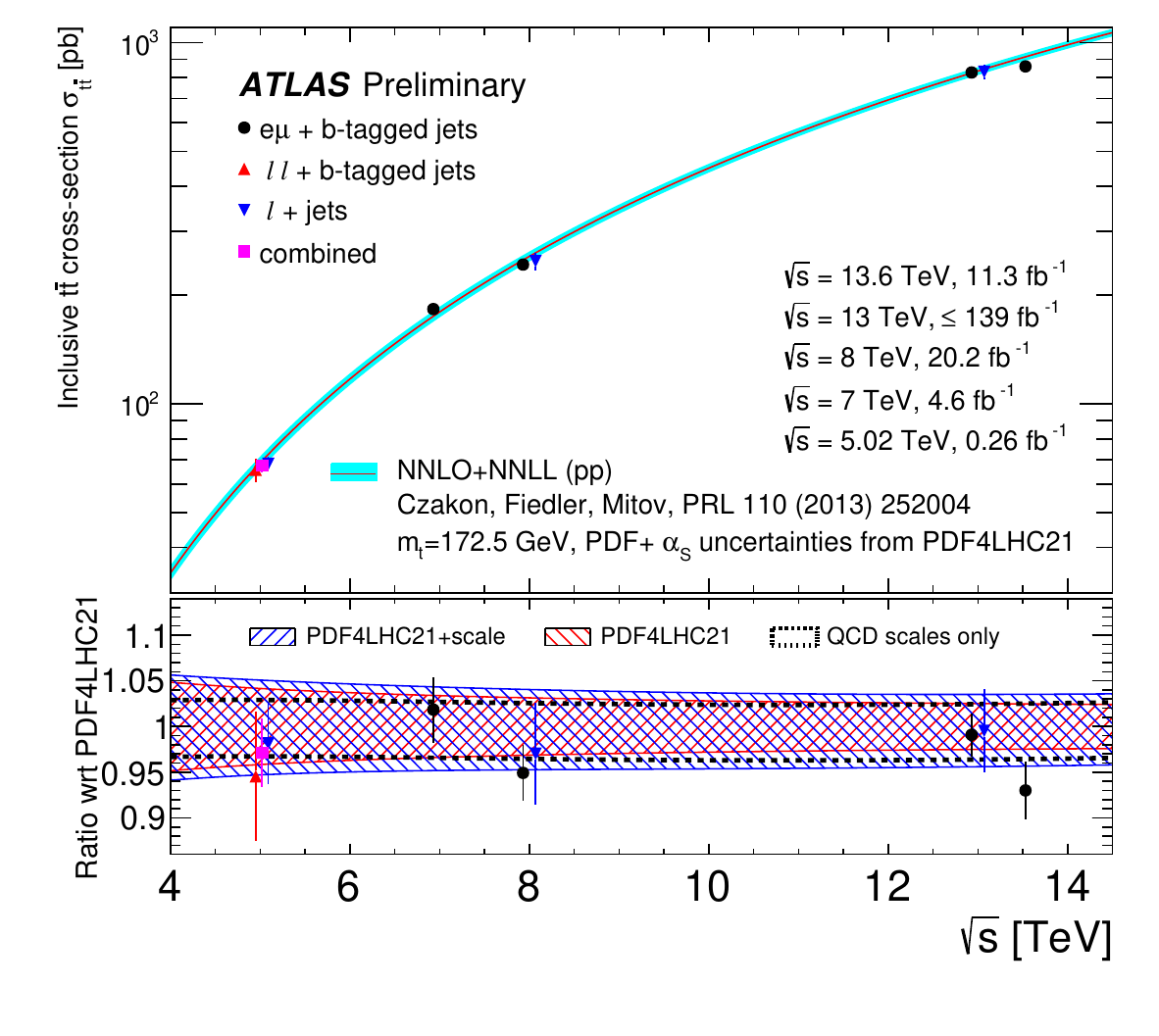}
			\caption{}
		\end{subfigure}
		\begin{subfigure}{0.48\textwidth}
			\includegraphics[scale=0.32]{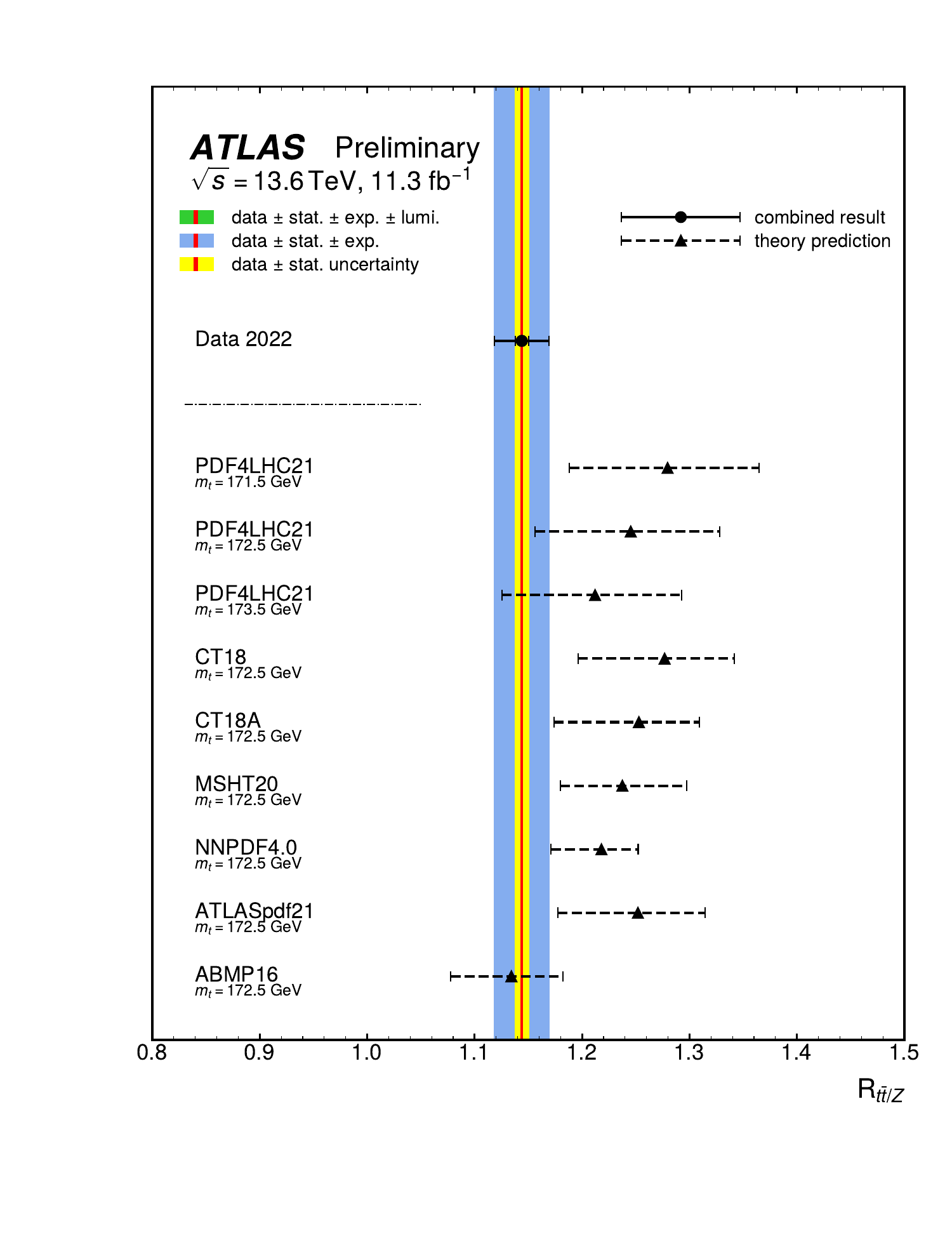}
			\caption{}
		\end{subfigure}
		\caption{(a) Comparison of the measured $t\bar{t}$ cross-sections at various centre-of-mass energies and the theory predictions using the PDF4LHC21 PDF set. The bottom panel shows the ratio of the measured values and three predictions that either contain only the uncertainties originating from the QCD scale variations (black), only the variations in the PDF uncertainties (red) or the total uncertainty in the prediction (blue). (b) Ratio of the \ttbar to the Z-boson cross-section compared to the prediction for several sets of parton distribution functions. For the PDF4LHC21 PDF set, predictions for different assumptions about the top-quark mass are also displayed.~\cite{ATLAS-CONF-2023-006}}
		\label{fig:ttbarRun3/xSec}
	\end{figure}

	The measurement of the $t\bar{t}$ and $Z$-boson cross-section and their ratio at $\sqrt{s}=13.6$ TeV denotes the first measurement using data from Run 3~\cite{ATLAS-CONF-2023-006}.
	The data used in this analysis amount to an integrated luminosity of $11.3$ fb$^{-1}$.
	Due to the early stage of data taking, many detector uncertainties are large, while ATLAS is deriving precise measurements of the luminosity and in-situ calibrations for leptons and jets.
	The measurement is already limited by systematic uncertainties, with luminosity uncertainties and lepton efficiency uncertainties being the largest sources.\\
	Events with an opposite-charged lepton pair and either one or two $b$-tagged jets are selected to measure the $t\bar{t}$ cross-section.
	The $Z$-boson production cross-section is measured in a fiducial space in $ee$ and $\mu\mu$ final states, where the invariant  mass of the leptons is required to be within the $Z$-boson mass window. 
	Additionally, a measurement of the ratio of the $t\bar{t}$ and the $Z$-boson production cross-sections is performed, which benefits from cancellation of several systematic uncertainties.
	Since $t\bar{t}$ and $Z$-boson production dynamics are driven to a large extent by different proton constituents at the LHC, the ratio of these cross-sections has a significant sensitivity to the gluon-to-quark PDF ratio.
	A profile likelihood fit is used to extract the $t\bar{t}$ and $Z$-boson cross-sections and - similarly to the measurement shown in Section~\ref{sec:tchannel} - combined jet selection and $b$-tagging efficiency.
	A second profile likelihood fit is performed afterwards in which the parameter of interest is the ratio $R_{t\bar{t}/Z}$, rather than the $t\bar{t}$ cross-section.
	This procedure ensures that all correlations among systematic uncertainties in the $t\bar{t}$ and $Z$-boson cross-section measurement are taken into account.
	In Figure~\ref{fig:ttbarRun3/dataMC}, the leading lepton $\pt$ distribution for both fiducial spaces are shown.
	Simulated events are in good agreement within the overall uncertainties in the low $\pt$ regions.
	This is important, as the lepton uncertainties impact the profile likelihood fit via their acceptance in this region.
	The results of this measurement are
	\begin{gather}
	\sigma_{t\bar{t}} = 859 \pm 4\, \left(\mathrm{stat.}\right) \pm 22\, \left(\mathrm{syst.}\right) \pm 19\, \left(\mathrm{lumi.}\right)\,\mathrm{pb}\\
	\sigma_{Z}^{\mathrm{fid.}} = 751 \pm 0.3\, \left(\mathrm{stat.}\right) \pm 15\, \left(\mathrm{syst.}\right) \pm 17\, \left(\mathrm{lumi.}\right)\,\mathrm{pb}\\
	R_{t\bar{t}/Z} = 1.114 \pm 0.006\, \left(\mathrm{stat.}\right) \pm 0.022\, \left(\mathrm{syst.}\right) \pm 0.003\, \left(\mathrm{lumi.}\right)
	\end{gather}
	As illustrated in Figure~\ref{fig:ttbarRun3/xSec}, the measured $t\bar{t}$ cross-section is very close but in agreement with the SM prediction using the PDF4LHC21 PDF set~\cite{Butterworth:2015oua}. The largest uncertainty for the \ttbar cross-section is the theory uncertainty on the modelling of the signal parton shower.

	\subsection{Measurement of total and differential $t\bar{t}W$ cross-section}
	\label{sec:ttW}
\begin{figure}
		\centering
		\includegraphics[scale=0.35]{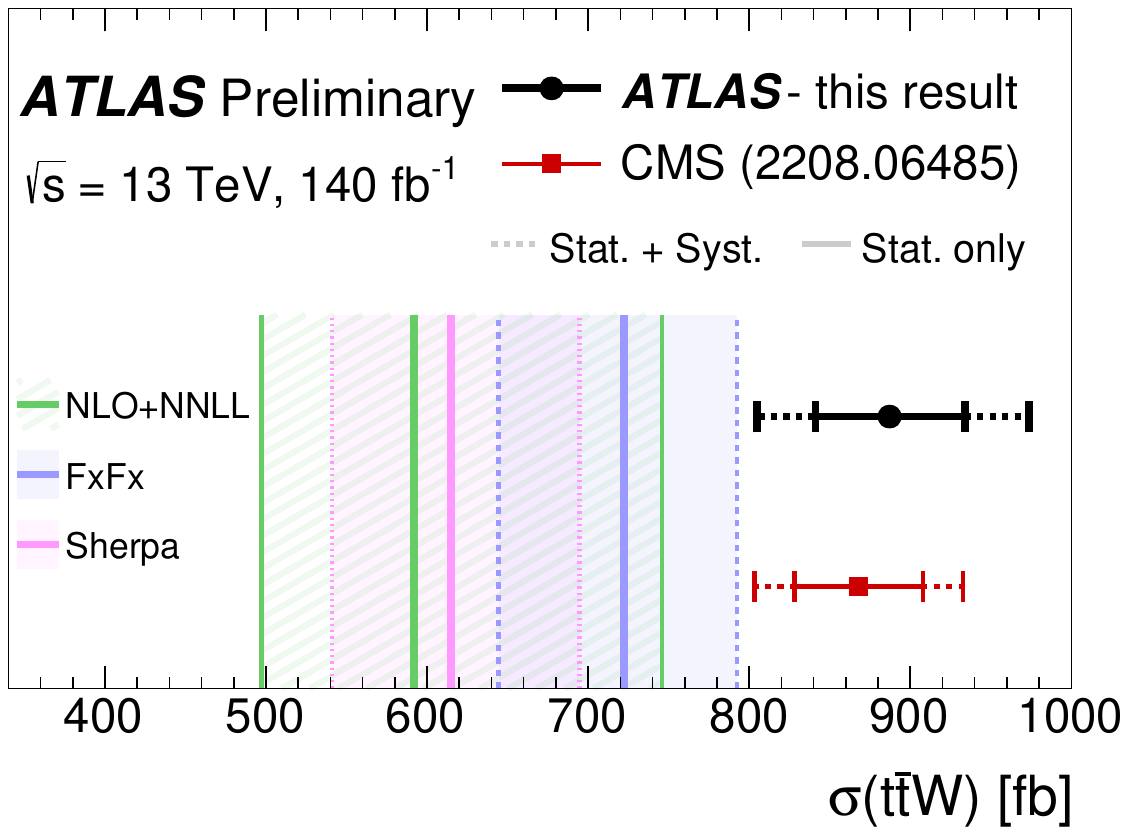}
		\caption{Comparison of the measured inclusive $t\bar{t}W$ cross-section to the theoretical predictions from \Sherpa, the \MCatNLO FxFx prescription including EWK corrections from Ref.~\cite{Frederix_2021}, the NLO+NNLL prediction from Ref.~\cite{Kulesza_2020} and the measurement from CMS~\cite{CMS-TOP-21-010}.~\cite{ATLAS-CONF-2023-019}}
		\label{fig:ttW/excess}
	\end{figure}

	The measurement of the total and differential $t\bar{t}W$ cross-section is a very important analysis, as this process is a large background for many BSM searches but also other Higgs- and top-quark measurements.
	Further, both, CMS and ATLAS, observed an excess in $t\bar{t}W$ events in many analyses.
	Using the full Run-2 dataset amounting to an integrated luminosity of $140$ fb$^{-1}$ allows for the first differential measurement of this process~\cite{ATLAS-CONF-2023-019}.
	Events are required to have exactly two or three leptons, at least two jets and at least one or two $b$-tagged jets.
	Thus, the main background processes are $t\bar{t}Z/\gamma$, diboson and $t\bar{t}H$ production.
	For these, as well as for backgrounds from charge mis-identified leptonts, control regions are assigned.
	A maximum likelihood fit is used to extract the cross-section of $t\bar{t}W$.
	The result, illustrated in Figure~\ref{fig:ttW/excess}, is within 1.5$\sigma$ with the SM prediction.
	\begin{equation}
	\sigma(t\bar{t}W) = 890 \pm 50 \left(\mathrm{stat.}\right) \pm 70 \left(\mathrm{syst.}\right)\,\mathrm{fb}
	\end{equation}
	The largest uncertainties are signal modelling uncertainties and prompt lepton uncertainties.
	Nevertheless, this result improves the relative uncertainty by more than a factor of two with respect to the previous analysis~\cite{TOPQ-2015-22}.
	CMS also has done a precise measurement of this process~\cite{CMS-TOP-21-011}, observing the same discrepancies between data and the MC prediction.\\
	$t\bar{t}W$ also allows to measure a production asymmetry for the signal process.
	In this measurement, the production asymmetry and its ratio, $ R(t\bar{t}W)$, are measured to be
	\begin{align}
	\sigma_{t\bar{t}W^+} &= 585^{+35}_{-34}\left(\mathrm{stat.}\right)^{+47}_{-44}\left(\mathrm{syst.}\right)\, \mathrm{fb}\\
	\sigma_{t\bar{t}W^-} &= 301^{+28}_{-27}\left(\mathrm{stat.}\right)^{+35}_{-31}\left(\mathrm{syst.}\right)\, \mathrm{fb}\\
	R(t\bar{t}W) &= 1.95^{+0.21}_{-0.18}\left(\mathrm{stat.}\right)^{+0.16}_{-0.13}\left(\mathrm{syst.}\right),
	\end{align}
	where the ratio $R(t\bar{t}W)$ is in good agreement with the prediction from \Sherpa~\cite{Gleisberg:2008ta}~\cite{Hoeche:2009rj}.\\
	Absolute and normalised differential cross-section measurements are performed.
	Profile-likelihood unfolding in seven observables was done.
	The differential measurement is limited by statistical uncertainties.
	Again, the absolut differential cross-sections are larger than the theoretical predictions, which is in agreement with the inclusive cross-section result.
	This shows, that future theoretical developments are needed (e.g. predictions at NNLO in QCD) to understand the discrepancy better.
	Also, the Run 3 dataset of the LHC will provide more data to further probe this final state.  
	
	\subsection{Observation of $t\bar{t}t\bar{t}$}
	\label{sec:4tops}
\begin{figure}
		\centering
		\includegraphics[scale=0.35]{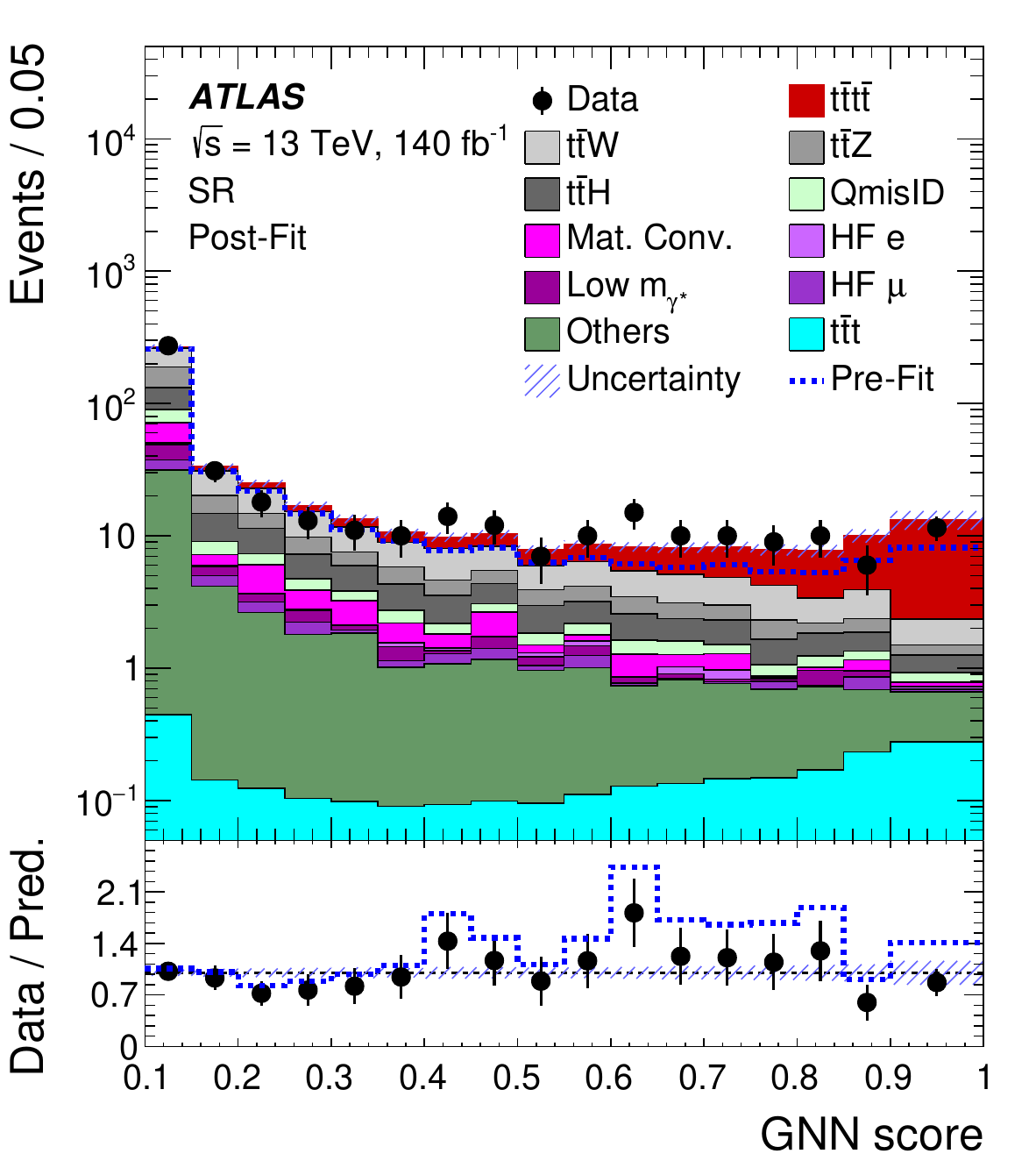}
		\caption{Comparison between data and the predictions after a fit to data for the GNN distribution in the signal region. The first bin contains underflow events. The ratio of the data to the total post-fit prediction is shown in the lower panel. The dashed blue lines show the pre-fit prediction in the upper panel and the ratio of the data to the total pre-fit prediction in the lower panel. The shaded band represents the total post-fit uncertainty in the prediction.~\cite{TOPQ-2021-08}}
		\label{fig:4tops/GNN}
	\end{figure}

	The observation of four-top-quark production ($t\bar{t}t\bar{t}$)~\cite{TOPQ-2021-08} is a very important measurement, as new particles or even new forces could alter the probability of this process.
	The predicted SM cross-section with $\sigma^{SM}_{t\bar{t}t\bar{t}}=13.4^{+1.0}_{-1.8}$ fb~\cite{vanbeekveld2022threshold} is very small, compared to the cross-section of \ttbar production.
	Many improvements were implemented with respect to the previous analysis~\cite{TOPQ-2018-05}, which found a strong evidence for this process.
	An improved particle identification allows for lower $\pt$ requirements, which then allow to select more events.
	Data-driven estimates for background processes as $t\bar{t}W$, but also for mis-identified or non-prompt leptons are introduced and also the treatment of the $t\bar{t}t$ background is improved.
	For this analysis, events with exactly two same-charge leptons or at least three leptons are selected.
	Additionally, events are required to have at least six jets, of which two need to be $b$-tagged.
	A Graph neural network (GNN) is used to distinguish the signal from the background.
	The GNN output distribution, which is used to extract the cross-section, is shown in Figure~\ref{fig:4tops/GNN}.
	
\begin{figure}
		\centering
		\includegraphics[scale=0.35]{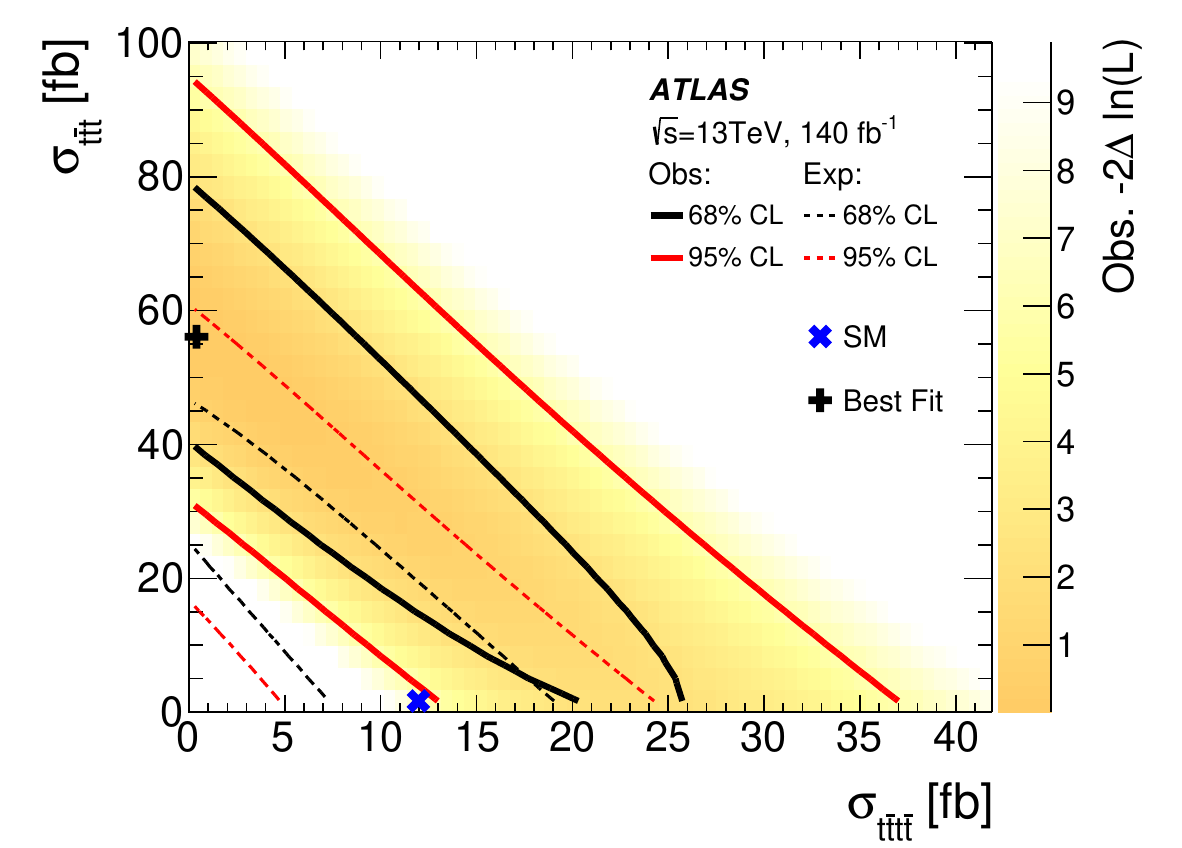}
		\caption{Two-dimensional negative log-likelihood contour for the $t\bar{t}t$ cross-section ($\sigma_{t\bar{t}t}$) versus the $t\bar{t}t\bar{t}$ cross-section ($\sigma_{t\bar{t}t\bar{t}}$) when the normalisation of both processes are treated as free parameters in the fit. The blue cross shows the SM expectation of $\sigma_{t\bar{t}t\bar{t}}$=12 fb and $\sigma_{t\bar{t}t}$=1.67 fb, both computed at NLO, while the black cross shows the best-fit value. The observed (expected) exclusion contours at 68\% (black) and 95\% CL (red) are shown in solid (dashed) lines. The gradient-shaded area represents the observed likelihood value as a function of $\sigma_{t\bar{t}t}$ and $\sigma_{t\bar{t}t\bar{t}}$.~\cite{TOPQ-2021-08}}
		\label{fig:4tops/ttt}
	\end{figure}

	A binned profile likelihood fit is used to determine the normalisation of the largest backgrounds and the cross-section.
	The result, $\sigma_{t\bar{t}t\bar{t}}=22.5^{+6.6}_{-5.5}$ fb, is observed with 6.1$\sigma$ significance. 
	It is 1.8$\sigma$ above the SM prediction~\cite{vanbeekveld2022threshold}.
	This measurement is further used to set limits on the cross-section of three top-quark production $t\bar{t}t$, which are illustrated in Figure~\ref{fig:4tops/ttt}.
	Three further interpretations were done, setting limits on four heavy flavour fermion EFT operators as well as on the top-quark Yukawa coupling and the Higgs oblique parameter.
	
	\subsection{Observation of single-top-qiark production with a photon}
	\label{sec:tqy}
\begin{figure}
		\centering
		\includegraphics[scale=0.35]{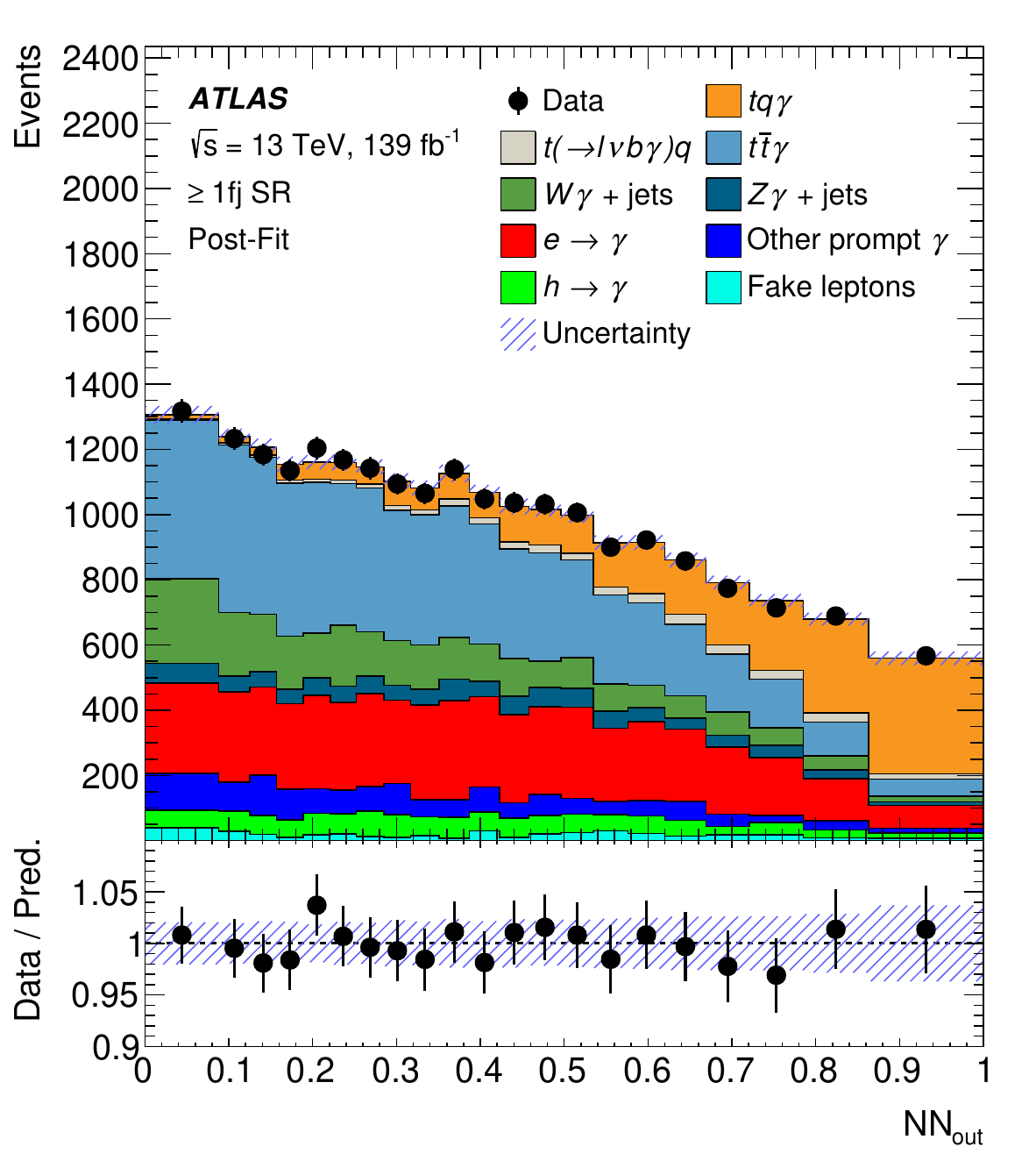}
		\caption{Distribution of the DNN output in the $\geq 1$ forward jet SR in data and the expected contribution of the signal and background processes after the profile-likelihood fit. The hashed band represents the uncertainties in the SM prediction.~\cite{ATLAS-CONF-2022-013}}
		\label{fig:tqy/DNN}
	\end{figure}

	Another rare process is observed in single-top-quark production together with a photon~\cite{ATLAS-CONF-2022-013}.
	Using the full Run-2 dataset, events were selected with exactly one photon, exactly one lepton, exactly one $b$-tagged jet and either zero or one forward jet.
	This forward jet is characteristic for the signal process.
	Photon fakes are a crucial background for this analysis and it is estimated using data-driven methods.
	The main backgrounds in this measurement are $t\bar{t}\gamma$ and $W\gamma$.
	A deep neural network (DNN) is used to separate the signal from the backgrounds.
	The output distribution is used in the profile likelihood fit, which is done to extract the cross-sections.
	The post-fit distribution of the DNN output is shown in Figure~\ref{fig:tqy/DNN}.
	Two fiducial cross-sections are measured, one on parton level, $\sigma_{tq\gamma} \times \mathcal{B}\left( t\rightarrow \ell\nu b\right)$ and one on particle level, $\sigma_{tq\gamma} \times \mathcal{B}\left( t \rightarrow \ell\nu b\right)$.
	The first one includes only events where the photon originates from the top-quark itself, while the latter one also includes photons which originate from the top-quark charged decay products.
	The measured cross-sections are:
	\begin{gather}
	\sigma_{tq\gamma} \times \mathcal{B}\left( t\rightarrow \ell\nu b\right) = 688 \pm 23\left(\mathrm{stat.}\right)^{+75}_{-71}\left(\mathrm{syst}\right) \mathrm{fb}\\
	\sigma_{tq\gamma} \times \mathcal{B}\left( t \rightarrow \ell\nu b\right) + \sigma_{t\left(\rightarrow\ell\nu b \gamma\right)q} = 303 \pm 9\left(\mathrm{stat}\right)^{+33}_{-32}\left(\mathrm{syst}\right) \mathrm{fb}
	\end{gather}
	This process is observed with an observed (expected) significance of 9.3$\sigma$ (6.8$\sigma$).
	The dominant uncertainties in this measurement arise from the signal modelling. 
	\FloatBarrier

	\subsection{Measurement of top-quark mass using a template method}
	\label{sec:topmass}
\begin{figure}
		\centering
		\includegraphics[scale=0.35]{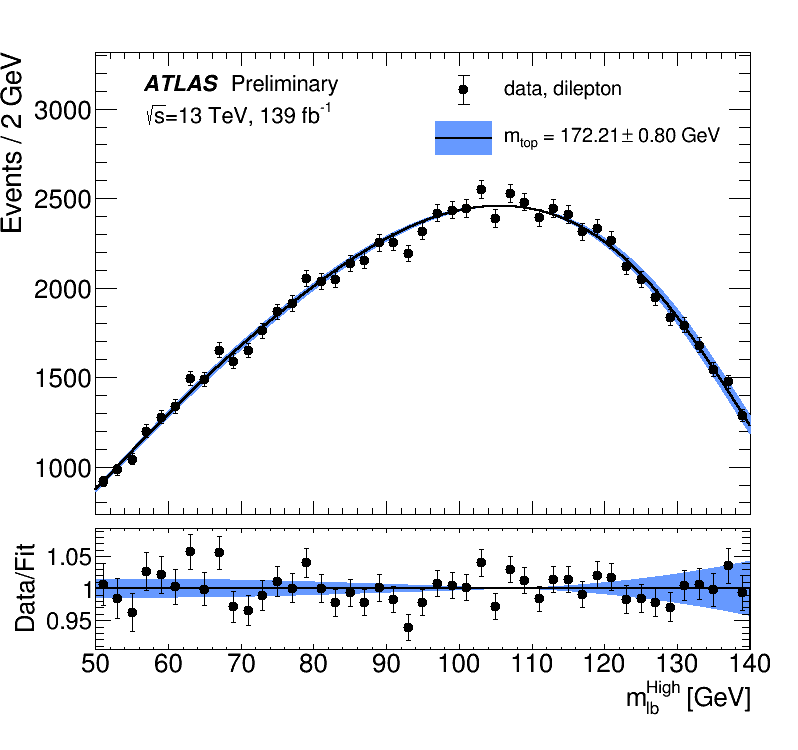}
		\caption{The High $m_{\ell b}$ distribution in data compared to the predicted distribution for the measured top-quark mass value. The data points and the template fit function to this data are shown in black. The blue uncertainty band is constructed by varying the template fit function within the full uncertainty of the measurement. The lower panel shows the ratio of data and the template fit function. The $\chi^2$/ndf of the best-fit result is 55.3/44 with a probability of $P(\chi, ndf)=0.24$.~\cite{ATLAS-CONF-2022-058}}
		\label{fig:topmass/template}
	\end{figure}

	The top-quark mass is measured using the full Run-2 dataset~\cite{ATLAS-CONF-2022-058}.
	Events with exactly two leptons and at least two jets are selected.
	Further, exactly two of the jets need to be $b$-tagged.
	A DNN is used to pair the $b$-jet and lepton from one top-quark decay, where only events with a DNN score $<0.6$ are selected.
	This improved method extracts the invariant mass $m_{\ell b}$, which is very sensitive to the top-quark mass, for each event and helps reducing modelling and jet related uncertainties.
	For the template method, distributions are constructed for a number of discrete values of the top-quark mass.
	These are interpolated, such that the final template function only depends on one free parameter, which represents the top-quark mass.
	An unbinned likelihood fit to data is done with this template function, where the fit range is optimised to minimise the total uncertainty.
	The post-fit template distribution is shown in Figure~\ref{fig:topmass/template}
	The top-quark mass is measured to be $m_{\mathrm{top}} = 172.21 \pm 0.20 (\mathrm{stat.}) \pm 0.67 (\mathrm{syst.}) \pm 0.39(\mathrm{recoil})$ GeV.
	Overall, signal modelling uncertainties are significantly improved, but a new systematic uncertainty needed to be introduced.
	The recoil uncertainty describes gluon radiation recoiling against the top-quark.
	New parton shower algorithm such as VINCIA are expected to mitigate this uncertainty in the future.
	
	\subsection{Evidence for charge asymmetry in $pp\rightarrow t\bar{t}$}
	\label{sec:ttcharge}
\begin{figure}
		\centering
		\includegraphics[scale=0.13]{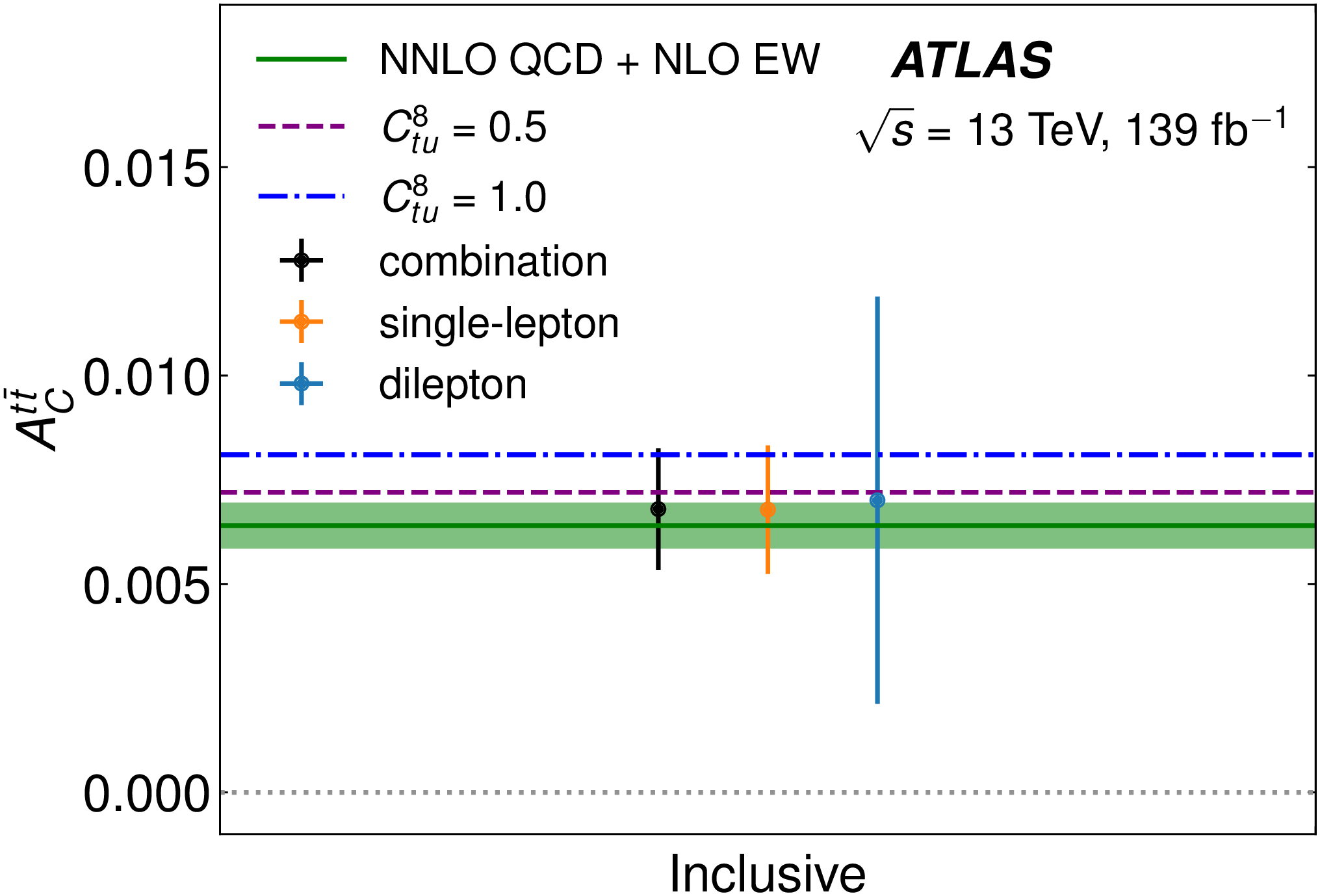}
		\caption{The unfolded inclusive charge asymmetriy. Vertical bars correspond to the total uncertainties. Shaded regions show SM theory predictions calculated at NNLO in QCD and NLO in EW theory.~\cite{TOPQ-2020-06}}
		\label{fig:ttcharge/result}
	\end{figure}

	Due to using proton-proton collision, a central-forward asymmetry is a very small effect at the LHC ($\mathcal{O}(1\%)$).
	In $t\bar{t}$ events, this asymmetry was measured using the difference between the absolute value of the top-quark rapidity, $\lvert y_t \rvert$, and the absolute value of the top-antiquark rapidity $\lvert y_{\bar{t}}$, to construct a charge asymmetry $A_C^{t\bar{t}}$~\cite{TOPQ-2020-06}:
	\begin{equation}
	A_{\mathrm{C}}^{t\bar{t}} = \frac{N\left( \Delta\lvert y_{t\bar{t}}\rvert>0\right)-N\left( \Delta\lvert y_{t\bar{t}}\rvert<0\right)}
	{N\left( \Delta\lvert y_{t\bar{t}}\rvert>0\right)+N\left( \Delta\lvert y_{t\bar{t}}\rvert<0\right)}
	\end{equation}
	Events with single or dilepton final states are targeted, as well as with high $\pt$ hadronic top-quark decays.
	Data-driven methods are used for fake lepton background and a BDT is used to match quarks to jets in order to enhance the reconstruction.
	To compare the data to a fixed-order theory prediction, fully bayesian unfolding is used.
	In Figure~\ref{fig:ttcharge/result}, the measured charge asymmetry in the single-lepton, dilepton and combined channel is shown.
	The measured charge asymmetry value is $A_C^{t\bar{t}}= 0.0068 \pm 0.0015 \left(\mathrm{stat.}+\mathrm{syst.}\right)$, which differs 4.7$\sigma$ from zero and thus gives a strong evidence for the charge asymmetry.
	The measurement is limited by the statistical uncertainties.
	
	
	\section{Conclusion}
	\label{sec:conclusion}
	
	The top-quark working group in ATLAS presented many new and improved measurements using the full Run-2 dataset amounting to an integrated luminosity of $\mathcal{L}=140$ fb$^{-1}$.
	The statistical precision of the full Run-2 dataset is exploited and thus, the top-quark properties, such as the top-quark mass, are measured with good precision.
	Accordingly, further rare processes (four-top-production, tq$\gamma$) are observed for the first time.
	Different types of machine-learning approaches and models are used in various ways in many analyses, e.g. to reconstruct the signal process or separate the signal and background processes.
	New and improved data-driven methods to determine background processes which are not modelled well enough in simulations, help to gain more precise results and a better understanding of these processes.
	As every measurement depends on well-simulated physics processes, differential cross-section measurements are required to help understanding MC generator predictions and also show that theoretical progress is needed, e.g. for $t\bar{t}W$ modelling.
	For the recoil from gluons directly from the top-quark, a new systematic uncertainty is taken into account.
	Most analyses are also limited by systematic uncertainties now.
	This is also expected for measurements using data from the ongoing LHC Run 3 using a centre-of-mass energy of $\sqrt{s}=13$ TeV.
	The first measurement using Run 3 data of $t\bar{t}$ and $Z$-boson cross-sections is also shown in this document.
	The results are already limited by systematic uncertainties in this early stage of data taking, and are in agreement with the SM prediction.
	In general, using the data obtained with the LHC Run 3 will allow for even higher precision measurements and for improving analyses, which are currently limited by data statistics, e.g. asymmetry measurements.\\
	In this document, only a very small fraction of recent results are shown.
	Many more results are already obtained and can be found here: \href{https://twiki.cern.ch/twiki/bin/view/AtlasPublic/TopPublicResults}{ATLAS Top Public results}.
	For LHC Run-2, but especially for Run 3, further improved measurements and even more precise results are yet to come.

	\printbibliography
	
	
	
\end{document}